\documentclass[%
 reprint,
superscriptaddress,
 amsmath,amssymb,
 aps,
]{revtex4-1}
\usepackage{amsmath}
\numberwithin{equation}{section}
\renewcommand{\theequation}{\arabic{section}.\arabic{equation}}

\usepackage{float}

\usepackage{graphicx}
\usepackage{dcolumn}
\usepackage{bm}
\usepackage{times}
\usepackage[colorlinks]{hyperref}
\hypersetup{
    citecolor=[rgb]{0,0,1},
    linkcolor=[rgb]{0,0,1},   
    urlcolor=[rgb]{0,0,0.9}}
\usepackage{wasysym}
\usepackage{mathtools} 

\usepackage{diagbox}

\usepackage{filecontents}

\immediate\write18{bibtex orbifoldQH}
\usepackage{natbib}

\begin{document}


\title{Coupled Wire Model of $Z_2 \times Z_2$ Orbifold Quantum Hall States}

\author{Pok Man Tam}
\affiliation{Department of Physics and Astronomy, University of Pennsylvania, Philadelphia, PA 19104, USA}
\author{Yichen Hu}
\affiliation{Department of Physics and Astronomy, University of Pennsylvania, Philadelphia, PA 19104, USA}
\affiliation{Rudolf Peierls Centre for Theoretical Physics, Clarendon Laboratory, Parks Road, Oxford, OX1 3PU, United Kingdom}
\author{Charles L. Kane}
\affiliation{Department of Physics and Astronomy, University of Pennsylvania, Philadelphia, PA 19104, USA}

\begin{abstract}
We construct a coupled wire model for a sequence of non-Abelian quantum Hall states occurring at filling factors $\nu=2/(2M+q)$ with integers $M$ and even(odd) integers $q$ for fermionic(bosonic) states. They are termed $Z_2 \times Z_2$ orbifold states, which have a topological order with a neutral sector described by the $c=1$ orbifold conformal field theory (CFT) at radius $R_{\rm orbifold}=\sqrt{p/2}$ with even integers $p$. When $p=2$, the state can be viewed as two decoupled layers of Moore-Read (MR) state, whose neutral sector is described by the Ising $\times$ Ising CFT and contains a $Z_2 \times Z_2$ fusion subalgebra. We demonstrate that orbifold states with $p>2$, also containing a $Z_2 \times Z_2$ fusion algebra, can be obtained by coupling together an array of MR$\times$MR wires through local interactions. The corresponding charge spectrum of quasiparticles is also examined. The orbifold states constructed here are complementary to the $Z_4$ orbifold states, whose neutral edge theory is described by orbifold CFT with odd integer $p$ and contains a $Z_4$ fusion algebra. 
\end{abstract}

\maketitle

\section{\label{sec1}Introduction}
The fractional quantum Hall effect provides a fertile setting for studying topological electronic states with richly structured patterns of quantum entanglement \citep{01}.   The low energy structure of a quantum Hall state is characterized by a 2+1D topological quantum field theory that characterizes the charge and statistics of its elementary quasiparticles \citep{02,03,04,05,06,07}. A closely related 2D conformal field theory (CFT) characterizes the edge states\citep{08,09,10,11,20}, where the primary fields of the CFT are associated with the quasiparticle types. A simple and explicit connection between the CFT and the TQFT is provided by the coupled wire construction \citep{8,9,3,1}, which takes strips of quantum Hall states with edges described by the 1+1D CFT, and couples them by electron tunneling to form a 2+1D topological phase.

The simplest Abelian fractional quantum Hall states are described by an Abelian 2+1D Chern-Simons theory, and have edge states that can be expressed in terms of free bosons and have an integer chiral central charge $c$ \citep{09}.  The connection to conformal field theory enabled Moore and Read \citep{10}, and later Read and Rezayi \citep{11} to envision non-Abelian states, which are a subject of intense interest because of their proposed application to quantum computation \citep{17,15,21}. Non-Abelian states feature non-Abelian anyons, along with topological ground state degeneracies. In general, their edge states have non-integer central charge, and can not be described in terms of free bosons compactified on a circle. 
 
\begin{figure}[t!]
   \includegraphics[width=9cm,height=6cm ]{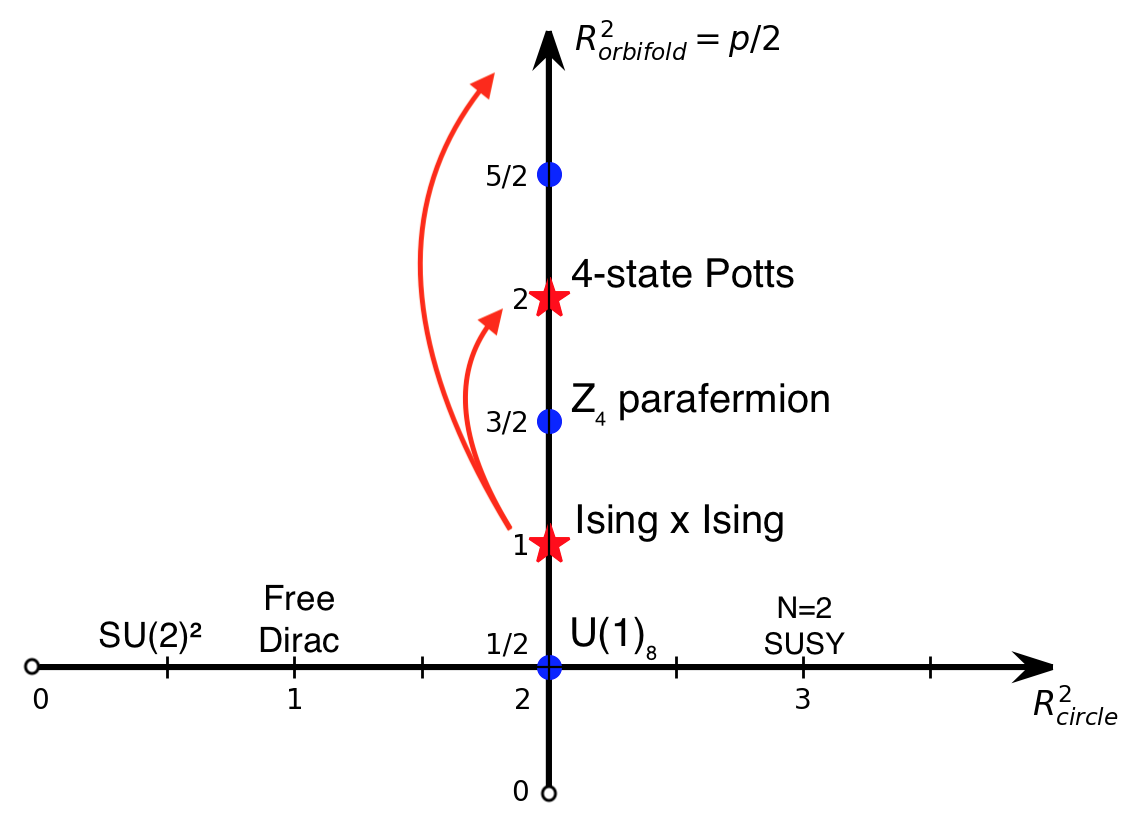}\centering
  \caption{\small{Moduli space of $c=1$ conformal field theory. Orbifold quantum Hall states correspond to a sequence of discrete points on the vertical orbifold line, with radius $R_{\rm orbifold}=\sqrt{p/2}$ for integer $p$. Coupled wire construction for the odd-$p$ states is provided by Ref. \citep{1}, and in this paper the even-$p$ states are constructed. With $p=2$ as the starting point, one can tune the orbifold radius, \textit{i.e.} change $p$, by local interaction. As we will see in Sec. \ref{sec5}, our construction requires $p$ to be an even integer in order to gap out the bulk neutral sector.}}
  \label{c=1CFT}
\end{figure}

In this paper we consider a class of non-Abelian quantum Hall states derived from the CFT of a free boson defined on an orbifold, that is a circle in which angles $\varphi$ and $-\varphi$ are identified. This type of CFT was studied extensively in the 1980's in the context of string theory \citep{2,4,12}. The free boson on an orbifold of radius $R_{\rm orbifold}$ and the free boson on a circle of radius $R_{\rm circle}$ define a continuous space of $c=1$ CFT's as illustrated in Fig. \ref{c=1CFT}. The two lines parametrizing the orbifold and circle theories intersect at a single point, $U(1)_8$, which can be described as either the circle at $R_{\rm circle} = \sqrt{2}$ or the orbifold at $R_{\rm orbifold} =1/ \sqrt{2}$. The orbifold CFT is almost the same as the circle CFT (which is equivalent to the theory of a conventional Luttinger liquid), except that it contains additional twist operators. 

Orbifold quantum Hall states were first studied by Barkeshli and Wen \citep{5,18,19}. There is a sense in which they form the simplest of the non-Abelian states. They possess non-Abelian quasiparticles with associated topological ground state degeneracies, but the edge states have integer central charge $c$. Using effective field theory, slave-particle formulation and the ``pattern of zeros" method, Barkeshli and Wen identified quantum Hall states associated with orbifolds with a discrete series of radii, $R_{\rm orbifold} = \sqrt{p/2}$, where $p$ is an integer. At these radii the orbifold CFT's were fully characterized by Dijkgraaf, Vafa, Verlinde and Verlinde \citep{2}, and it was shown that the two cases in which the integer $p$ is even or odd have different structures to their fusion algebras. The case in which $p$ is odd, which includes the special cases of $U(1)_8$ ($p=1$) and $Z_4$ parafermions ($p=3$), have a quasiparticle with a $Z_4$ fusion algebra, in which a cluster of four identical quasiparticles can fuse to the identity. In contrast, when $p$ is even, which includes the special cases of Ising $\times$ Ising ($p=2$) and 4-state Potts ($p=4$), there is a $Z_2\times Z_2$ structure, which includes two sets of quasiparticles for which a pair of identical quasiparticles fuse to the identity.

In a previous paper \citep{1}, a coupled wire construction was introduced to show that the orbifold states with odd integer $p$ arise naturally in a theory in which electrons cluster into groups of four.  These states form a generalization of the $Z_4$ parafermion Read-Rezayi state, whose wavefunction can also be interpreted in terms of the clustering of 4 electrons. However, this wire construction did not describe the states with even integer $p$, and it was clear that the $4e$-clustering was inconsistent with the $Z_2\times Z_2$ fusion algebra of the even-$p$ orbifold theory. If one instead consider the $2e$-clustering of electrons, that is to couple together Luttinger liquids in the paired state, a $Z_2$ fusion algebra can indeed arise. Such a construction has led to the Moore-Read (MR) state as discussed in Ref. \citep{3}.

\begin{figure}[t!]
   \includegraphics[width=9.5cm,height=3.3cm ]{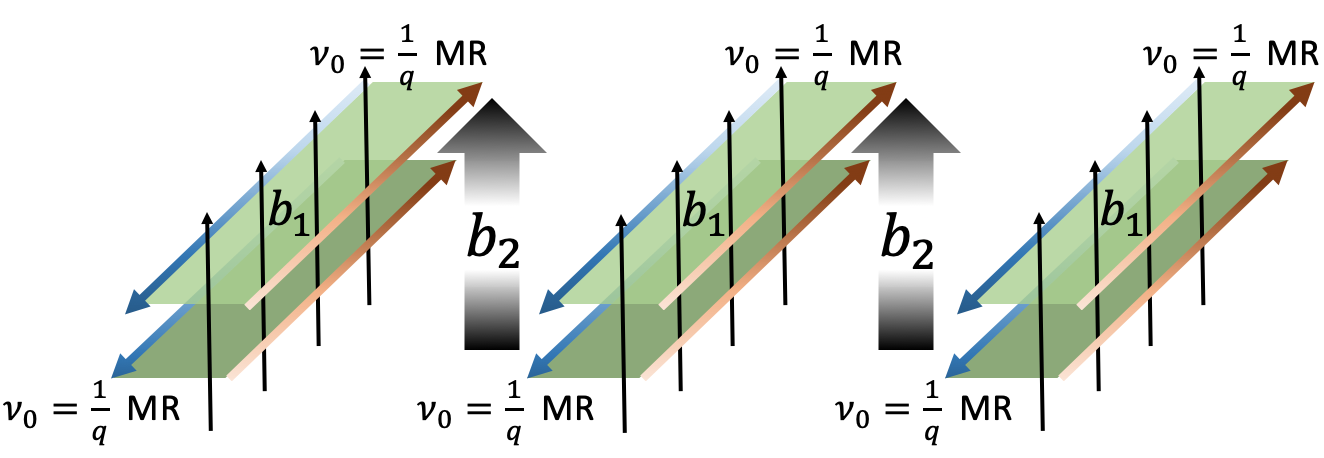}\centering
  \caption{\small{Coupled wire construction for the $Z_2 \times Z_2$ orbifold quantum Hall states. Each wire is composed of two identical layers of Moore-Read Pfaffian states at filling $\nu_0=1/q$. The model requires flux insertion both within each wire and between wires, with the density of the former denoted as $b_1$ and the latter denoted as $b_2$. }}
  \label{wiremodel}
\end{figure}

In this paper, we introduce a coupled wire construction for the even-$p$ orbifold states following the above reasoning. Our starting point, as depicted in Fig. \ref{wiremodel}, is a set of strips consisting of two layers of identical MR states, which are each described by the Ising TQFT. When the strips are glued together in the trivial way, a bilayer MR state is formed, whose quasiparticle structure in the neutral sector is described by the Ising $\times$ Ising theory, which can be identified with the $p=2$ point of the orbifold theory. We show that, by introducing local interactions within and between the strips, we can construct the sequence of orbifold quantum Hall states described by all even integers $p$. We initiate to name them as the $Z_2 \times Z_2$ orbifold states so as to reflect their fusion structure.

The rest of the paper is organized as follows. In Sec. \ref{sec3}, we provide a detailed study of a single wire of MR$\times$MR quantum Hall strip, first explaining how the orbifold theory arises in the neutral sector (Sec.\ref{sec3.1.0}), as well as the associated fusion algebra that interests us (Sec.\ref{sec3.1.1}). We then systematically construct physical operators in our theory and present their bosonized representation (Sec.\ref{sec3.1.2}), which immediately allow us to demonstrate how the orbifold radius in a single wire can be tuned by local intra-wire interaction (Sec.\ref{sec3.2}). Proceeding with an array of such wires in Sec. \ref{sec4}, we gap out the bulk charge sector by introducing charge-$2e$ inter-wire tunneling which leads to a bilayer Halperin state for strongly paired electrons. In Sec. \ref{sec5}, we further introduce charge-$e$ tunneling interaction to gap the bulk neutral sector, which leaves a pair of gapless chiral edges described by the orbifold CFT at $R_{\rm orbifold} = \sqrt{p/2}$ with even integer $p$. The completely gapped bulk would be described by a non-Abelian TQFT as requried by the bulk-boundary correspondence. The quasiparticle spectrum of the even-$p$ orbifold quantum Hall states, which is specific to the construction being presented here, is analyzed in Sec.\ref{sec7}. Concluding remarks, together with comparisons to the orbifold states proposed by Barkeshli and Wen, would be given in Sec. \ref{sec6}. 

\section{\label{sec3}Single Wire}

\subsection{\label{sec3.1}Moore-Read $\times$ Moore-Read}
A single wire in our model is a quantum Hall strip comprised of two layers (to be labeld by $\uparrow,\downarrow$ respectively) of Moore-Read Pfaffian state at filling factor $\nu_0=1/q$, with underlying electrons being fermionic (bosonic) when integer $q$ is even (odd). A coupled wire construction of a single layer of MR state has been discussed in Ref.\citep{3}. As is well-known, the edge CFT of Moore-Read Pfaffian state consists of a $c=1$ $U(1)$ chiral boson theory in the charge sector and a $c=1/2$ critical Ising theory in the neutral sector, so that the edge Hamiltonian of a single layer of MR can be written as 
\begin{equation}
\mathcal{H}_{\text{MR}} = \mathcal{H}^{\rho}_{\text{MR}}+\mathcal{H}^{\sigma}_{\text{MR}}
\end{equation}
with
\begin{subequations}
\begin{align}
\mathcal{H}^{\rho}_{\text{MR}} &=  \frac{v_\rho}{2\pi} \sum_{r=R,L} (\partial_x \phi^\rho_r)^2, \\ 
\mathcal{H}^\sigma_{\text{MR}} &= iv_F (\gamma_R \partial_x \gamma_R - \gamma_L \partial_x \gamma_L).
\end{align}
\end{subequations}
Here $\phi^\rho_{r}$ is the chiral bosonic charge mode, which is compactified on a circle such that $\phi^\rho_{r} \equiv \phi^\rho_{r} +2\pi$, and $\gamma_{R/L}$ is the chiral Majorana neutral mode. With two layers of MR in a single wire, the charge sector is simply composed of two decoupled $U(1)$ chiral boson theories, with the Hamiltonian
\begin{equation}\label{chargeMRHamiltonian}
\begin{split}
\mathcal{H}^{\rho}_j &= \frac{v_\rho}{2\pi} \sum_{a=\uparrow,\downarrow}[(\partial_x \phi^{\rho,a}_{j,R})^2+(\partial_x \phi^{\rho,a}_{j,L})^2]\\
&=  \frac{v_\rho}{\pi} \sum_{a=\uparrow,\downarrow}[(\partial_x \varphi^{\rho,a}_{j})^2+(\partial_x \theta^{\rho,a}_{j})^2].
\end{split}
\end{equation}
Here $j$ labels the wire and $a=\uparrow/\downarrow$ labels the layer. We also introduce the conjugate bosonic variables $\{\varphi,\theta\}$, which are related to the chiral fields by $\phi_{R/L} = \varphi \pm \theta$. The chiral charge fields obey the following commutation relation,
\begin{equation}\label{oldbosoncommutation}
[\phi^{\rho,a}_{j,r}(x),\phi^{\rho,b}_{j',r'}(x')] = iq\pi r\delta_{r,r'}\delta_{j,j'}\delta_{a,b}\text{ sgn}(x-x')
\end{equation}
with $a,b=\uparrow/\downarrow$ and $r,r'=R/L=+1/-1$.

As for the neutral sector, one may simply view it as two decoupled critical Ising models. Alternatively, we want to take a more interesting perspective, by viewing the Ising $\times$ Ising CFT as a $c=1$ orbifold conformal field theory at radius $R_{\rm orbifold}=1$ \citep{2,4}. This perspective motivates a generalization of the Moore-Read $\times$ Moore-Read quantum Hall state to a family of $Z_2 \times Z_2$ orbifold states, as we will explain below.

\subsubsection{\label{sec3.1.0} Orbifold theory in neutral sector}

Combining two Majorana fermions into a Dirac fermion such that $\psi_{j,r} = \gamma^\uparrow_{j,r}+i\gamma^\downarrow_{j,r}$ (with $r=R,L$), the neutral sector Hamiltonian of the bilayer system can be written as
\begin{equation}\label{majoranahamiltonian}
\begin{split}
\mathcal{H}^\sigma_j &= iv_F \sum_{\substack{a=\uparrow,\downarrow}} (\gamma^a_{j,R} \partial_x \gamma^a_{j,R}-\gamma^a_{j,L} \partial_x \gamma^a_{j,L}) \\
& = iv_F (\psi_{j,R}^\dagger \partial_x \psi_{j,R}-\psi_{j,L}^\dagger \partial_x \psi_{j,L}).
\end{split}
\end{equation}
It is important to note that, since the Ising model is dual to a Majorana fermion coupled to a dynamical $Z_2$ gauge field \citep{7}, we have to take into account the effects of two copies of $Z_2$ gauge symmetry here, which leads to redundant labeling of physical states. To explore the consequences, we bosonize the complex fermion as
\begin{equation}
\psi_{j,r} = \gamma^{\uparrow}_{j,r} + i\gamma^{\downarrow}_{j,r} \sim e^{i\phi^{\sigma}_{j,r}}.
\end{equation}
where $\phi^\sigma_{j,r}$ is the chiral bosonic neutral mode. Applying $Z_2$ symmetry on both layers, which identifies $\psi_{j,r}$ with $-\psi_{j,r}$, we obtain the shift symmetry
\begin{equation}\label{radius1/2}
\phi^\sigma_{j,r} \mapsto \phi^\sigma_{j,r} + \pi.
\end{equation}
Applying $Z_2$ symmetry on a single layer, which identifies $\psi_{j,r}$ with $\psi_{j,r}^\dagger$, we obtain the orbifold symmetry
 \begin{equation}
 \phi^\sigma_{j,r} \mapsto -\phi^\sigma_{j,r}.
  \end{equation} 
Equivalently, the conjugate bosonic fields have the following symmetry transformations,
 \begin{subequations}\label{bosonsym}
 \begin{align}
 &\varphi^\sigma_{j} \mapsto \varphi^\sigma_{j} +\pi, \quad\theta^\sigma_{j} \mapsto \theta^\sigma_{j}+2\pi.\\
 &\varphi^\sigma_{j} \mapsto -\varphi^\sigma_{j}, \quad\quad\theta^\sigma_{j} \mapsto -\theta^\sigma_{j}.
 \end{align}
  \end{subequations} 
In contrast to the $\pi$-shift symmetry of $\varphi^\sigma$ as effected by the $Z_2$ gauge symmetry, $\theta^\sigma$ retains its usual $2\pi$-shift symmetry since it is not affected in (\ref{radius1/2}). The above transformation thus defines an orbifold theory of circle with radius $R_{\rm orbifold}=1$ \citep{13}. The corresponding Hamiltonian, bosonized from (\ref{majoranahamiltonian}), takes the form
\begin{equation}\label{neutralMRHamiltonian}
\begin{split}
\mathcal{H}^{\sigma}_j &= \frac{v_\sigma}{2\pi}[(\partial_x \phi^{\sigma}_{j,R})^2+(\partial_x \phi^{\sigma}_{j,L})^2]\\
&=  \frac{v_\sigma}{\pi}[(\partial_x \varphi^{\sigma}_{j})^2+(\partial_x \theta^{\sigma}_{j})^2].
\end{split}
\end{equation}

It will be useful for latter discussion to mention that the chiral neutral fields are defined to obey the following commutation relation,
\begin{equation}\label{sigmacommutation}
[\phi^{\sigma}_{j,r}(x),\phi^\sigma_{j',r'}(x')] = i\pi \text{ sgn}(x_{j,r} - x_{j',r'}).
\end{equation}
Particularly, the coordinates here are ordered in a ``raster pattern", in which $... < x_{j,L} < x_{j,R} < x_{j+1,L} < x_{j+1,R} < ...$, by defining $x_{j,R}= L+x+2Lj$ and $x_{j,L} = L-x+2Lj$. Doing so, we can always ensure fermions on different wires to anticommute without the use of Klein factors. 

To sum up, a single MR $\times$ MR wire is described by the following Hamiltonian
\begin{equation}
\mathcal{H}_j = \mathcal{H}^{\rho}_{j}+\mathcal{H}^{\sigma}_{j}
\end{equation}
with the charge and neutral sectors defined in (\ref{chargeMRHamiltonian}) and (\ref{neutralMRHamiltonian}) respectively. This is the starting point of our coupled wire construction. We will describe below the bosonization of local operators, such as the creation/annihilation operator for edge electrons and operators that scatter quasiparticles within the MR layer, so as to prepare ourselves for the study of allowed interactions between wires. But before that, let us first investigate into the orbifold theory just established. After all, our goal is to construct quantum Hall states whose neutral edge modes are described by such theories. The orbifold theories have non-trivial operator product structures, which translate into exotic fusions of non-Abelian quasiparticles in the bulk.

\subsubsection{\label{sec3.1.1}$Z_2 \times Z_2\;$ fusion algebra}
Let us first understand better our starting point, where there are two decoupled Ising models in the neutral sector. For a single Ising model, we denote the vacuum, Majorana fermion and Ising spin operator as $\textbf{1}, \gamma$ and $\sigma$ respectively. The Ising $\times$ Ising primary fields are summarized in Table \ref{Ising2}. The table also shows a correspondence, at the operator level, between Ising $\times$ Ising model and the $c=1$ orbifold conformal field theory at radius $R_{\rm orbifold}=1$.  The latter model at a generic radius $R_{\rm orbifold}=\sqrt{p/2}$ ($p\in \mathbb{Z}$), also known as the $Z_2$ orbifold of $Z_{2p}$ Gaussian model, is first studied by Dijkgraaff \textit{et al.} in Ref.\citep{2}.

\begin{table}[t!]
\centering
\begin{tabular}{ c | c | c  }
Notation in Ref.\citep{2} & Dimension & Operator ($p=2$) \\
 \hline
$1$ & $0$ & $\textbf{1}^{\uparrow} \textbf{1}^{\downarrow}$\\
$j$ & $1$ &  $\gamma^{\uparrow} \gamma^{\downarrow}$\\
$\phi^i_p\;(i=1,2)$ & $p/4$ &  $\textbf{1}^{\uparrow} \gamma^{\downarrow},\;\gamma^{\uparrow}\textbf{1}^{\downarrow}$\\
$\phi_k\;(k=1,...,p-1)$ & $k^2/4p$ &  $\sigma^{\uparrow}\sigma^{\downarrow}$\\
$\sigma_i\;(i=1,2)$ & $1/16$ &  $\textbf{1}^{\uparrow} \sigma^{\downarrow},\;\sigma^{\uparrow}\textbf{1}^{\downarrow}$\\
$\tau_i\;(i=1,2)$ & $9/16$ &  $\gamma^{\uparrow} \sigma^{\downarrow},\;\sigma^{\uparrow}\gamma^{\downarrow}$\\
\end{tabular}
\caption{Primary fields of the orbifold conformal field theory at radius $R_{\rm orbifold}=\sqrt{p/2}$. The $p=2$ case, which corresponds to two decoupled Ising models in the neutral sector of MR $\times$ MR, is illustrated explicitly on the right.}
\label{Ising2}
\end{table}

Following the well-known Ising fusion rules, we can easily check for $p=2$ that
\begin{subequations}\label{Z2fusion}
\begin{align}
j \times j &=1, \\
\phi^i_p \times \phi^i_p &=1, \\
\phi^1_p \times \phi^2_p &= j,
\end{align}
\end{subequations}
with $i=1,2$. Therefore, the operators $\{1, j, \phi^1_p, \phi^2_p\}$ form a $Z_2 \times Z_2$ fusion algebra. This turns out to be a general property of the $Z_2$ orbifold of $Z_{2p}$ Gaussian model for all even integers $p$. The non-Abelian nature of the theory can be seen by fusing the twist fields, which generate all the vertex operators through
\begin{subequations}\label{nonAbelianfusion}
\begin{align}
\sigma_i \times \sigma_i &= 1+\phi^i_p+\sum_{\mathclap{k\in \text{ even}}} \phi_k,\\
\sigma_1 \times \sigma_2 &= \sum_{\mathclap{k\in \text{ odd}}} \phi_k,\\
j \times \sigma_i &= \tau_i.
\end{align}
\end{subequations}
For a more elaborate discussion on the fusion algebra, we refer the interested readers to the standard reference \citep{2}.

A family of non-Abelian fractional quantum Hall states with such edge theories are first proposed by Barkeshli and Wen from the effective field theory perspective, and are termed the orbifold quantum Hall states \citep{5}. In this paper, we will present a coupled wire construction for these states, so as to provide support to their (theoretical) existence at a more microscopic level. Based on the general property of fusion algebra in this family, we initiate to name them as the $Z_2 \times Z_2$ orbifold states, in order to distinguish them from the $Z_4$ orbifold states, which are associated to $Z_2$ orbifold of $Z_{2p}$ Gaussian model for odd integer $p$ and have been constructed in a coupled wire model already \citep{1}.

\subsubsection{\label{sec3.1.2}Local operators and their bosonization}
In order to couple together an array of MR $\times$ MR wires to create a gapped bulk, and for the gapless edges to be described by an orbifold theory at an arbitrary even integer $p$, we have to turn on interactions within and between wires. In principle, the allowed interactions (also known as local interactions) are those that can be written as products of fundamental electron operators. In a quantum Hall state, these include operators that create/annihilate edge electrons and those that scatter quasiparticles from one edge to another. The quasiparticle spectrum of the Moore-Read Pfaffian state is well-known \citep{10} and summarized in Table \ref{MRspectrum}. This table will guide us to construct the useful local operators for our coupled wire construction.

We make use of two types of local operators: (1) ones that create/annihilate edge electrons, and change the charge on the edge by an integer; (2) ones that scatter fractionally charged quasiparticles from one edge to another. For the first case, we have the local electron operator,
\begin{equation}\label{edgeelectron}
\Psi_{e,r}^a \sim \gamma^a_r e^{i\phi^{\rho,a}_r}
\end{equation}
with $a=\uparrow/\downarrow$ and $r=R/L=+1/-1$ (for simplicity, we suppress the wire label in this subsection, as all operators introduced here correspond to the same wire). As discussed earlier in Sec. \ref{sec3.1.0}, by combining Majorana fermions from different layers into a Dirac fermion we can bosonize as follows,
\begin{subequations}
\begin{align}
\gamma^\uparrow_{r} &\sim \cos\phi^\sigma_{r} = \cos(\varphi^\sigma+r\theta^\sigma),\\
\gamma^\downarrow_{r} &\sim \sin\phi^\sigma_{r} = \sin(\varphi^\sigma+r\theta^\sigma),
\end{align}
\end{subequations}
where the anticommutation of fermions is ensured by  (\ref{sigmacommutation}). Applying (\ref{edgeelectron}) twice, one can annihilate a charge-$2e$ pair locally. Due to the fusion $\gamma \times \gamma =\textbf{1}$, the corresponding operator is trivial in the neutral sector,
\begin{equation}\label{bare2e}
\Phi_{2e,r}^a \sim e^{2i\phi^{\rho,a}_r}.
\end{equation}

\begin{table}[t!]
\centering
\begin{tabular}{c | c c c c c}
\diagbox{neutral}{charge} &\;\;\;0\;\;\;&\;\;\;$e/4$\;\;\;&\;\;\;$e/2$\;\;\;&\;\;\;$3e/4$\;\;\;&\;\;\;$e$\;\;\;\\ 
\hline
\textbf{1}& $\Circle$ & & $\odot$ & & $\odot$\\
$\sigma$& &$\odot$ & &$\odot$ &\\
$\gamma$& $\odot$ & & $\odot$ & & $\CIRCLE$\\
\end{tabular}
\caption{Particle spectrum of the Moore-Read Pfaffian state at filling factor $\nu_0=1/2$. The symbols $\odot$ indicate existing quasiparticles, $\CIRCLE$ indicates the electron and $\Circle$ indicates the vacuum. The generalized Moore-Read state at filling factor $\nu_0=1/q$ has a similar spectrum, which particularly contains the charge-$e/q$ Abelian quasiparticle and non-Abelian $\sigma$-particle of charge-$e/2q$. Local operators can be constructed by considering scattering processes of the above particles.}
\label{MRspectrum}
\end{table}

Next, we look at local operators that scatter quasiparticles in the MR state at $\nu_0=1/q$. One that is trivial in the neutral sector describes backscattering of charge-$e/q$ Abelian quasiparticles, 
\begin{equation}
\mathcal{V}_{1}^a\sim (e^{\frac{i}{q}\phi^{\rho,a}_L})^\dagger e^{\frac{i}{q}\phi^{\rho,a}_R} \sim e^{\frac{2i}{q}\theta^{\rho,a}}.
\end{equation}
By repeatedly applying $\mathcal{V}^{\uparrow/\downarrow}_{1}$, we further obtain
\begin{equation}\label{e/2scattering}
(\mathcal{V}_{1}^{\uparrow})^l (\mathcal{V}_{1}^{\downarrow})^k \sim e^{\frac{2i}{q}(l\theta^{\rho,\uparrow}+k\theta^{\rho,\downarrow})},
\end{equation}
which is local as long as $l$ and $k$ are integers.

Local operators that are trivial in the charge sector can be constructed by scattering neutral fermions. The following two operators will be of particular importance to our wire construction,
\begin{subequations}
\begin{align}
\mathcal{V}_{2\gamma} &= i(\gamma^{\uparrow}_R \gamma^{\uparrow}_L + \gamma^{\downarrow}_R \gamma^{\downarrow}_L) \sim \cos 2 \theta^\sigma,\\
\mathcal{V}_{4\gamma} &=  \gamma^{\uparrow}_R \gamma^{\downarrow}_R \gamma^{\uparrow}_L \gamma^{\downarrow}_L \sim \partial_x \phi^\sigma_R\;\partial_x \phi^\sigma_L.
\end{align}
\end{subequations}
Finally, we will also make use of the operator that backscatters a pair of charge-$e/2q$ non-Abelian $\sigma$-particles (one from each layer), which is non-trivial in both the charge and neutral sector,
\begin{equation}\label{4sigma}
\mathcal{V}_{4\sigma} = \sigma^{\uparrow}_{R}\sigma^{\uparrow}_{L}\sigma^{\downarrow}_{R}\sigma^{\downarrow}_{L} e^{\frac{i}{q}(\theta^{\rho,\uparrow}+\theta^{\rho,\downarrow})}.
\end{equation}
Though it is not straightforward to bosonize the neutral sector, it should be noted that $\sigma^{\uparrow}_{R}\sigma^{\uparrow}_{L}\sigma^{\downarrow}_{R}\sigma^{\downarrow}_{L}$ acts as a Jordan-Wigner string to the Dirac fermion $\psi_{R/L}$\citep{3}. Thus its bosonized representation should contain only $e^{in\theta^\sigma}$ with odd integers $n$. In addition, for the local scattering interaction to be invariant under the orbifold symmetry $\theta^\sigma \rightarrow -\theta^\sigma$, the bosonized representation must be a linear combination of $\cos n \theta^{\sigma}$. Therefore, we can write 
\begin{equation}\label{4sigmabosonized}
\mathcal{V}_{4\sigma} \sim  \Big(\sum_{n\in \text{ odd}} B_n\cos n\theta^{\sigma}\Big)e^{\frac{i}{q}(\theta^{\rho,\uparrow}+\theta^{\rho,\downarrow})}.
\end{equation}
The coefficients $B_n$ assume generic values, as the scattering of $\sigma$-particles can always be accompanied by scatterings of neutral fermions, through the $\mathcal{V}_{2\gamma}$ term.

\subsection{\label{sec3.2}Changing the orbifold radius}
With local operators and their bosonized representations established for a single MR $\times$ MR wire, we now demonstrate how the orbifold radius in the neutral sector can be adjusted, so as to obtain generalizations of the MR $\times$ MR quantum Hall states.  We include the forward-scattering of Majorana fermions $\mathcal{V}_{4\gamma}$, so that the neutral sector Hamiltonian becomes
\begin{equation}
\bar{\mathcal{H}}^{\sigma}_j = \frac{v_\sigma}{2\pi}[(\partial_x \phi^{\sigma}_{j,R})^2+2\kappa(\partial_x \phi^\sigma_{j,R})(\partial_x \phi^\sigma_{j,L})+ (\partial_x \phi^{\sigma}_{j,L})^2].
\end{equation}
Adjusting the coupling $\kappa$ such that $p=2\sqrt{(1-\kappa)/(1+\kappa)}$ and introducing a new set of chiral fields,
\begin{subequations}\label{neutralchiralfields}
\begin{align}
\bar{\phi}^{\sigma}_{j,R} &= \sqrt{\frac{2}{p}}\varphi^\sigma_j + \sqrt{\frac{p}{2}}\theta^\sigma_j,\\
\bar{\phi}^{\sigma}_{j,L} &= \sqrt{\frac{2}{p}}\varphi^\sigma_j - \sqrt{\frac{p}{2}}\theta^\sigma_j,
\end{align}
\end{subequations}
the Hamiltonian can then be re-diagonalized as
\begin{equation}\label{neutralpHamiltonian}
\begin{split}
\bar{\mathcal{H}}^{\sigma}_j &= \frac{\bar{v}_\sigma}{2\pi}[(\partial_x \bar{\phi}^{\sigma}_{j,R})^2+(\partial_x \bar{\phi}^{\sigma}_{j,L})^2]\\
&=\frac{\bar{v}_\sigma}{\pi}[\frac{2}{p}(\partial_x \varphi^{\sigma}_{j})^2+\frac{p}{2}(\partial_x \theta^{\sigma}_{j})^2]
\end{split}
\end{equation}
with $\bar{v}_\sigma = v_\sigma \sqrt{1-\kappa^2}$. Comparing with the $R_{\rm orbifold}=1$ theory in (\ref{neutralMRHamiltonian}), it shows that the orbifold radius has been rescaled to $R_{\rm orbifold}=\sqrt{p/2}$. Therefore, starting from a MR $\times$ MR wire and turning on the local interaction $\mathcal{V}_{4\gamma}$, $R_{\rm orbifold}$ can indeed be adjusted, allowing the theory to move along the orbifold line as depicted in Fig. \ref{c=1CFT}.

At this stage, one can in principle choose any value for $\kappa$ and then obtain an orbifold theory with an arbitrary radius. However, as we want to construct quantum Hall states with a gapped bulk, locality of inter-wire interaction would require $p/2$ to be an integer. This will be clear when we engineer inter-wire couplings to gap the neutral sector in Sec. \ref{sec5}. But before dealing with the more subtle issues there, let us first gap out the charge sector for an array of MR $\times$ MR wires, from which we can determine the sequence of filling fractions where these orbifold quantum Hall states would arise.\\

\begin{figure}[b!]
   \includegraphics[width=9cm,height=5cm ]{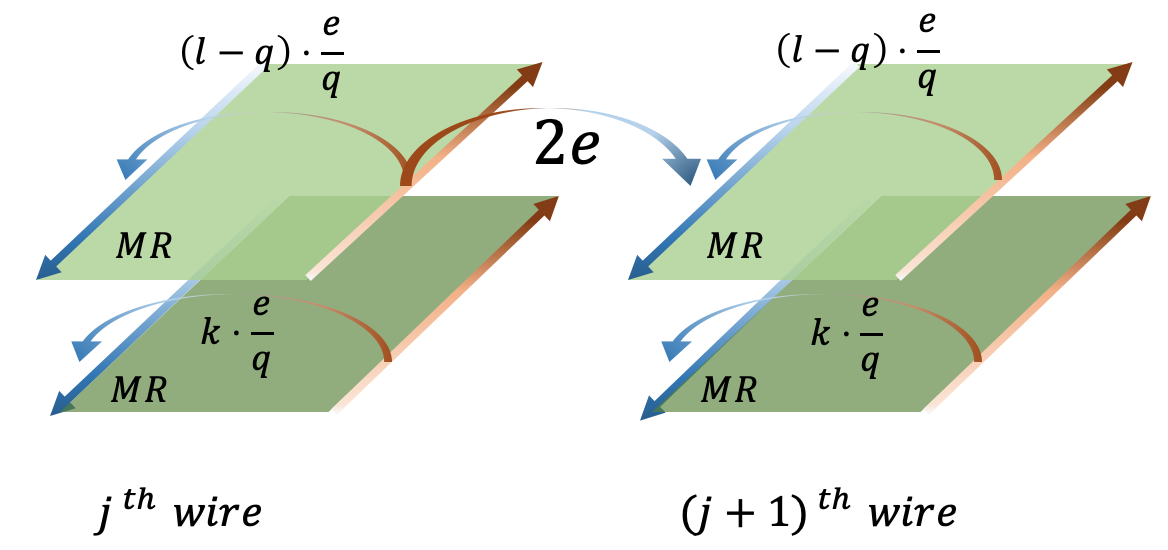}\centering
  \caption{\small{Illustration of the charge-$2e$ tunneling operator $( \Phi_{j+1,-l -k}^{\rho,\uparrow})^{\dagger} \Phi_{j,l k}^{\rho,\uparrow}$, which tunnels a pair of electrons across two neighboring wires through the top layer. This is associated with multiple intra-wire scatterings of charge-$e/q$ Abelian quasiparticles, as specified by $l$ and $k$. To completely gap out the charge sector, which has two components $\uparrow$ and $\downarrow$, a similar operator that tunnels a $2e$-pair through the bottom layer is also required.}}
  \label{2etunnelterm}
\end{figure}

\section{\label{sec4}Coupled Wires: Charge Sector}

We will now consider an array of MR $\times$ MR wires (Fig. \ref{wiremodel}) coupled together by local interactions of electron tunneling and quasiparticle scattering. In this section we focus on the two charge sectors (labeled by $\uparrow,\downarrow)$, and illustrate how they can be gapped by inter-wire tunneling of pairs of electrons, accompanied by intra-wire scattering of charge-$e/q$ Abelian quasiparticles. We begin by explicitly constructing such operators.\\ 

\subsection{\label{sec4.1}Charge-$2e$ tunneling operator}
Combining the bare charge-$2e$ operator (\ref{bare2e}) with backscatterings of charge-$e/q$ Abelian quasiparticles (\ref{e/2scattering}), we can define a local composite charge-$2e$ operator (on the $j$-th wire) for each layer
\begin{equation}
\begin{split}
\Phi_{j,l k}^{\rho,\uparrow} &\sim e^{2i(\varphi^{\rho,\uparrow}_{j}+\frac{l}{q}\theta^{\rho,\uparrow}_j+\frac{k}{q}\theta^{\rho,\downarrow}_j)},\\
\Phi_{j,l k}^{\rho,\downarrow} &\sim e^{2i(\varphi^{\rho,\downarrow}_{j}+\frac{l}{q}\theta^{\rho,\downarrow}_j+\frac{k}{q}\theta^{\rho,\uparrow}_j)},
\end{split}
\end{equation}
where $l,k \in \mathbb{Z}$. Here $l$ and $k$ are integers to ensure locality, as the number of scattering of Abelian quasiparticles in each layer is restricted to be an integer. To tunnel charge-$2e$ particles through the top or bottom layer, we consider the following terms
\begin{equation}
\mathcal{H}_{2,j+1/2}^{\uparrow/\downarrow} \sim  ( \Phi_{j+1,-l -k}^{\rho,\uparrow/\downarrow})^{\dagger} \Phi_{j,l k}^{\rho,\uparrow/\downarrow}+h.c.
\end{equation}
where for simplicity $l, k$ are chosen to be the same for both layers, so that our construction is symmetric under layer-exchange. The tunneling and scattering processes involved in the above operator are shown explicitly in Fig. \ref{2etunnelterm}. Upon bosonization, they can be written as
\begin{widetext}
\begin{subequations}
\begin{align}
\mathcal{H}_{2,j+1/2}^{\uparrow} &=-t_2^{\uparrow} \cos 2[ \varphi_j^{\rho,\uparrow}-\varphi_{j+1}^{\rho,\uparrow}+\frac{l}{q}(\theta_j^{\rho,\uparrow}+\theta_{j+1}^{\rho,\uparrow})+\frac{k}{q}(\theta_j^{\rho,\downarrow}+\theta_{j+1}^{\rho,\downarrow})],\\
\mathcal{H}_{2,j+1/2}^{\downarrow} &=-t_2^{\downarrow}\cos 2[ \varphi_j^{\rho,\downarrow}-\varphi_{j+1}^{\rho,\downarrow}+\frac{l}{q}(\theta_j^{\rho,\downarrow}+\theta_{j+1}^{\rho,\downarrow})+\frac{k}{q}(\theta_j^{\rho,\uparrow}+\theta_{j+1}^{\rho,\uparrow})].
\end{align}
\end{subequations}
\end{widetext}

The above expressions motivate us to introduce a new set of chiral bosonic fields
\begin{subequations}\label{newchargechiral}
\begin{align}
\bar{\phi}_{j,R}^{\rho,\uparrow} &= 2(\varphi_j^{\rho,\uparrow}+\frac{l}{q}\theta_{j}^{\rho,\uparrow}+\frac{k}{q}\theta_j^{\rho,\downarrow}),\\
\bar{\phi}_{j,L}^{\rho,\uparrow} &= 2(\varphi_j^{\rho,\uparrow}-\frac{l}{q}\theta_{j}^{\rho,\uparrow}-\frac{k}{q}\theta_j^{\rho,\downarrow}),\\
\bar{\phi}_{j,R}^{\rho,\downarrow} &= 2(\varphi_j^{\rho,\downarrow}+\frac{l}{q}\theta_{j}^{\rho,\downarrow}+\frac{k}{q}\theta_j^{\rho,\uparrow}),\\
\bar{\phi}_{j,L}^{\rho,\downarrow} &= 2(\varphi_j^{\rho,\downarrow}-\frac{l}{q}\theta_{j}^{\rho,\downarrow}-\frac{k}{q}\theta_j^{\rho,\uparrow}).
\end{align}
\end{subequations}
As such, the charge-$2e$ operator becomes $e^{i\bar{\phi}_{j,R}^{\rho,\uparrow/\downarrow}}$ and the tunneling terms can be written as $\cos (\bar{\phi}_{j,R}^{\rho,\uparrow}-\bar{\phi}_{j+1,L}^{\rho,\uparrow})$ and $\cos (\bar{\phi}_{j,R}^{\rho,\downarrow}-\bar{\phi}_{j+1,L}^{\rho,\downarrow})$ respectively. Following the spirit of wire construction, we anticipate a gapped bulk when these tunnelings flow to strong coupling. According to (\ref{oldbosoncommutation}), the new chiral fields obey \citep{6}
\begin{equation}\label{Kmatrix1}
[\partial_x \bar{\phi}^{\rho,a}_{j,r}(x), \bar{\phi}^{\rho,b}_{j',r'}(x')] = 2\pi i rK_{ab} \delta_{r,r'}\delta_{j,j'}\delta_{x,x'}
\end{equation}
where $a,b=\uparrow/\downarrow$, $r=R/L=+1/-1$ and the $K$-matrix is
\begin{equation}\label{Kmatrix2}
K_{ab} = 4\begin{pmatrix}
\;\; l \; & \;k \;\;\; \\ \;\;k\; & \;l\;\;\;
\end{pmatrix}.
\end{equation}
The charge density on the $j$-th wire is 
\begin{equation}\label{Kmatrix3}
\rho_j =\frac{1}{2\pi} \sum_{a=\uparrow,\downarrow} t_a \partial_x(\bar{\phi}^{\rho,a}_{j,R}-\bar{\phi}^{\rho,a}_{j,L}),
\end{equation}
with
\begin{equation}\label{Kmatrix4}
t_a = \frac{1}{2(l+k)} \begin{pmatrix}
1 \\ 1
\end{pmatrix}.
\end{equation}

For an array of wires, it is more convenient to consider variables associated to the links. For the $(j+1/2)$-link that connects the $j$-th and the $(j+1)$-th wire, we define
\begin{subequations}
\begin{align}
\tilde{\theta}_{j+1/2}^{\rho,a} &= (\bar{\phi}^{\rho,a}_{j,R} - \bar{\phi}^{\rho,a}_{j+1,L})/2,\\
\tilde{\varphi}_{j+1/2}^{\rho,a} &= (\bar{\phi}^{\rho,a}_{j,R} + \bar{\phi}^{\rho,a}_{j+1,L})/2,
\end{align}
\end{subequations}
which satisfy
\begin{equation}
\begin{split}
[\tilde{\theta}_{j+1/2}^{\rho,a}, \tilde{\theta}_{j'+1/2}^{\rho,b}] =[\tilde{\varphi}_{j+1/2}^{\rho,a} , \tilde{\varphi}_{j'+1/2}^{\rho,b} ]=0, \\
[\partial_x \tilde{\theta}_{j+1/2}^{\rho,a}, \tilde{\varphi}_{j'+1/2}^{\rho,b}] = i\pi K_{ab} \delta_{j,j'}\delta_{x,x'}.
\end{split}
\end{equation}
In terms of these link variables, $2e$-tunneling terms are expressed as
\begin{equation}\label{2etunnel}
\mathcal{H}^{a}_{2, j+1/2}= -t_2^{a} \cos 2\tilde{\theta}^{\rho,a}_{j+1/2}\quad (a=\uparrow,\downarrow).
\end{equation}
When $t_2^{\uparrow}$ and $t_2^{\downarrow}$ both flow to strong coupling, $\tilde{\theta}^{\rho,\uparrow}$ and $\tilde{\theta}^{\rho,\downarrow}$ will be locked simultaneously (as they commute) at integer multiples of $\pi$, and as such, the two bulk charge sectors will be gapped. Below, we will demonstrate how the tunneling terms can be ensured to flow to strong coupling, by incorporating a carefully designed inter-wire interaction in our model.

\subsection{\label{sec4.2}Gapping the charge sector}
Let us begin with the charge sector Hamiltonian in (\ref{chargeMRHamiltonian}). We include additional density-density scattering of the form $\partial_x \phi^{\rho,a}_R \partial_x \phi^{\rho,b}_L$, so that the Hamiltonian can be diagonalized by the new set of chiral fields $\bar{\phi}^{\rho,a}_{R/L}$ defined in (\ref{newchargechiral}),
\begin{equation}
\bar{\mathcal{H}}^{\rho}_j = \frac{\bar{v}_\rho}{2\pi}\sum_{a=\uparrow,\downarrow}[(\partial_x \bar{\phi}_{j,R}^{\rho,a})^2+(\partial_x \bar{\phi}_{j,L}^{\rho,a})^2].
\end{equation}
Setting up an array of $N$ such wires and before turning on any inter-wire tunneling, the bulk Hamiltonian can be written in terms of link variables as
\begin{equation}
\bar{\mathcal{H}}^{\rho}_{0,\text{bulk}} =\frac{\bar{v}_\rho}{\pi} \sum_{j=1}^{N-1} \sum_{a=\uparrow,\downarrow} [(\partial_x\tilde{\theta}^{\rho,a}_{j+1/2})^2+(\partial_x\tilde{\varphi}^{\rho,a}_{j+1/2})^2].
\end{equation}
At this stage, the charge $2e$-tunnelings described by (\ref{2etunnel}) are not guaranteed to gap the charge sector. As one can easily see for the simpler case when $k=0$, the scaling dimension $\Delta_a$ of $\cos 2\tilde{\theta}^{\rho,a}_{j+1/2}$ is
\begin{equation}
\Delta_a = 4l, 
\end{equation}
which implies that $t_2^a$ is perturbatively irrelevant for $l>0$.

In order to make $t_2^a$ relevant for generic $l$ and $k$, let us incorporate an inter-wire forward-scattering interaction so that the bulk Hamiltonian becomes
 \begin{equation}\label{chargeinterwire}
 \begin{split}
\tilde{\mathcal{H}}^{\rho}_{0,\text{bulk}} =\frac{\bar{v}_\rho}{2\pi} \sum_{j=1}^{N-1} \sum_{a=\uparrow,\downarrow}[(\partial_x \bar{\phi}^{\rho,a}_{j,R})^2 
&+ 2\lambda_\rho(\partial_x\bar{\phi}^{\rho,a}_{j,R})(\partial_x \bar{\phi}^{\rho,a}_{j+1,L})\\ &+ (\partial_x \bar{\phi}^{\rho,a}_{j+1,L})^2],
\end{split}
 \end{equation}
with $\lambda_\rho$ controlling the strength of forward-scattering. Writing in terms of link variables again, we obtain
\begin{equation}\label{linkHamiltonian}
\tilde{\mathcal{H}}^{\rho}_{0,\text{bulk}} =\frac{\tilde{v}_\rho}{\pi} \sum_{j=1}^{N-1} \sum_{a=\uparrow,\downarrow} [\frac{1}{g_\rho}(\partial_x\tilde{\theta}^{\rho,a}_{j+1/2})^2+g_\rho(\partial_x\tilde{\varphi}^{\rho,a}_{j+1/2})^2],
\end{equation}
where $g_\rho =\sqrt{(1+\lambda_\rho)/(1-\lambda_\rho)}$ and $\tilde{v}_\rho = \bar{v}_\rho \sqrt{1-\lambda_\rho^2}$. It is then clear that tuning $\lambda_\rho \rightarrow -1^-$ will make $g_\rho \rightarrow 0$ and ensure that $\Delta_{\uparrow/\downarrow} < 2 $ for arbitrary $l$ and $k$. Consequently, $t_2^{\uparrow}$ and $t_2^{\downarrow}$ all flow to strong coupling, and the array of wires with the bulk charge sector described by
\begin{equation}
\mathcal{H}^{\rho}_{\text{bulk}} = \tilde{\mathcal{H}}^{\rho}_{0,\text{bulk}} + \sum_{j=1}^{N-1} \sum_{a=\uparrow,\downarrow} \mathcal{H}_{2, j+1/2}^a
\end{equation}
will be gapped. The link variables $\tilde{\theta}^{\rho\uparrow}$ and $\tilde{\theta}^{\rho\downarrow}$ are then pinned at integer multiples of $\pi$.

In fact, if the neutral sector is gapped in a trivial way, namely the neutral sector in each wire is gapped individually by an intra-wire scattering of Majorana fermions (\textit{i.e.} $\mathcal{V}_{2\gamma}$), the fate of the coupled wire model would be an Abelian bilayer quantum Hall state for strongly-paired electrons, characterized by the $K$-matrix in (\ref{Kmatrix2}). Interestingly, as we will discuss in Sec. \ref{sec5}, there is alternative way to gap out the bulk neutral sector by tunneling a composite electron across neighboring wires, producing the non-Abelian $Z_2 \times Z_2$ orbifold quantum Hall states. 

\subsection{\label{sec4.3}Filling Fraction}
By specifying the tunneling operators that gap the charge sector, namely fixing the values of $l$ and $k$, the filling fraction is fixed accordingly. The most direct approach is to apply the $K$-matrix formalism \citep{14}. Using (\ref{Kmatrix1}-\ref{Kmatrix4}), we obtain \citep{6}
\begin{equation}
\nu= \textbf{t}^T K\;\textbf{t} = \frac{2}{l+k}.
\end{equation}
Given the microscopic picture provided by the coupled wire model, the above result can also be obtained from a more physical perspective. As explained below, the filling fraction is determined by requiring tunneling and scattering processes to altogether conserve momentum in the presence of background magnetic fields.

Recall that our starting point is a single wire of MR $\times$ MR bilayer quantum Hall state (each layer at $\nu_0=1/q$), whose implementation already requires certain flux insertion. Let us denote the one-dimensional (along wire) flux density inserted within the wire as $b_{1}$ and that inserted between wires as $b_2$. Then, the change of momentum (with the effect of Lorentz force taken into account) when a right-moving edge electron is scattered to a left-moving one within the wire is $b_1$, and when the electron is tunneled to the next wire the change is $b_2$. Here, the tunneling term required to gap the bulk is
\begin{equation}
 ( \Phi_{j+1,-l -k}^{\rho,\uparrow/\downarrow})^{\dagger} \Phi_{j,l k}^{\rho,\uparrow/\downarrow},
\end{equation}
which is depicted in Fig. \ref{2etunnelterm} and consists of the following processes: two right-moving edge electrons from the $j$-th wire are tunneled to the $(j+1)\text{-th}$ wire and become left-moving, and a total of $(k+l-q)$ charge-$e/q$ Abelian quasiparticles are scattered from right to left within each wire. Balancing the change of momenta in the above processes, we obtain
\begin{equation}
b_2=\frac{b_1}{q}(k+l-q).
\end{equation}
The filling fraction of this coupled wire construction is once again found to be
\begin{equation}\label{fillingfraction1}
\nu = \frac{2b_1/q}{b_1+b_2} =\frac{2}{k+l},
\end{equation} 
where we have used the fact that electron number density in a single wire is $b_1/q$ per Moore-Read layer. 

Finally, as we shall see in the next section, gapping the neutral sector further demands $l$ and $k$ to have the same (opposite) parity when $q$ is even (odd), so the orbifold states constructed in this paper actually correspond to the following sequence of filling fractions,
\begin{equation}\label{fillingfraction2}
\nu= \frac{2}{2M+q}
\end{equation}
with $M$ being an integer. This sequence only depends on the fermionic/bosonic nature of the orbifold state. Particularly, as we shall see next, it is independent of the orbifold radius. 

\section{\label{sec5}Coupled Wires: Neutral Sector}
With the charge sector gapped, we now proceed to gap the neutral sector. We begin with an array of wires whose neutral sector is described by the Hamiltonian in (\ref{neutralpHamiltonian}), \textit{i.e.} with the orbifold radius already tuned to $R_{\rm orbifold}=\sqrt{p/2}$ after incorporating appropriate intra-wire scattering of neutral fermions. We have to assume $p/2$ to be an integer in order to construct local operators, to be discussed below, that gap the bulk neutral sector. At the end, we will be left with a gapless chiral neutral mode at each edge characterized by the orbifold theory. 

\subsection{Charge-$e$ tunneling operator}

\begin{figure}[t!]
   \includegraphics[width=9cm,height=5cm ]{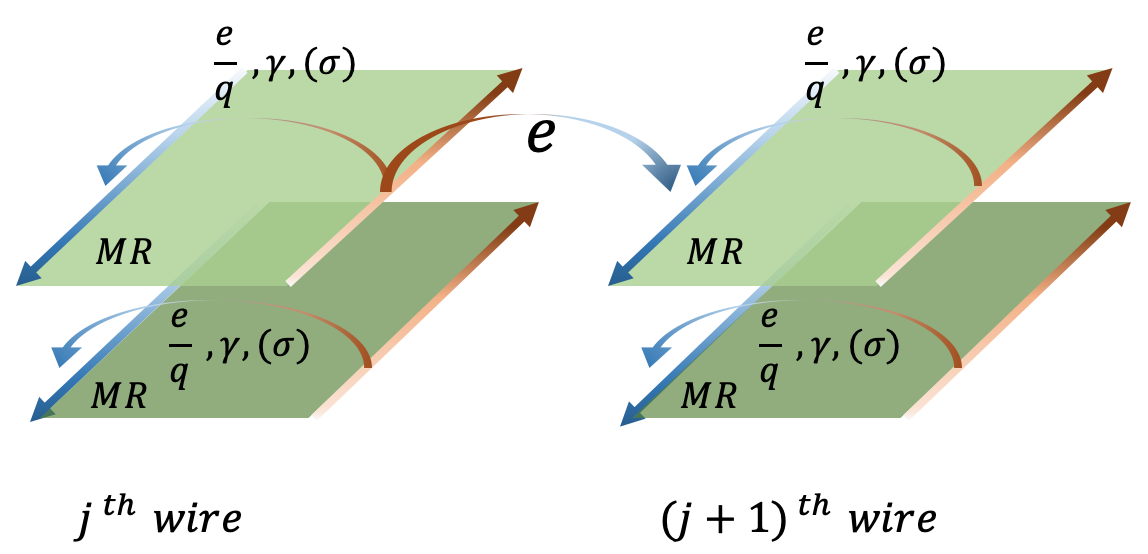}\centering
  \caption{\small{Illustration of the charge-$e$ tunneling operator $\Psi_{j+1,L}^{\dagger}\Psi_{j,R}$, which tunnels an edge electron across two neighboring wires through the top layer. This is associated with multiple intra-wire scatterings of charge-$e/q$ Abelian quasiparticles (specified by $l$ and $k$) and neutral Majorana fermions (by $\mathcal{V}_{2\gamma}$). When $p = 0$ mod $4$, additional intra-wire scatterings of charge-$e/2q$ $\sigma$-particles (by $\mathcal{V}_{4\sigma}$) are also required.}}
  \label{tunnelterm}
\end{figure}

To achieve our goal we need to construct a tunneling term of the form $\cos{\sqrt{\frac{p}{2}}(\bar{\phi}^{\sigma}_{j,R}-\bar{\phi}^{\sigma}_{j+1,L})}$, and ensure that this term flows to strong coupling. Depending on the parity of $p/2$, there are two distinct approaches to tunnel a single electron across neighboring wires in order to do this. A schematic of the tunneling process is provided in Fig. \ref{tunnelterm}. We first discuss the case when $p/2$ is odd, in which the associated tunneling term is simpler in the sense that intra-wire scattering of non-Abelian $\sigma$-particles is absent.
\subsubsection{p = 2 mod 4}
Let us consider tunneling the following composite electron
\begin{equation}\label{oddelectron}
\Psi_{j,r} \sim \Big(\sum_{\mathclap{n\in\text{ even}}} A_n\cos n\theta^{\sigma}_j \Big) \gamma^{\uparrow}_{j,r} e^{i(\varphi^{\rho,\uparrow}_{j}+r\frac{l}{q}\theta^{\rho,\uparrow}_j+r\frac{k}{q}\theta^{\rho,\downarrow}_j)}
\end{equation}
with $r=R/L=+1/-1$. The composite electron is formed by dressing the bare edge-electron $\Psi_{e,r}^\uparrow$ with $\cos n\theta^{\sigma}$ terms ($n\in 2\mathbb{Z}$), which are generated from the scattering of neutral Majorana fermions $\mathcal{V}_{2\gamma}$, introduced in Sec. \ref{sec3.1.2}. These terms are local by themselves without the charge sector. The electron is also dressed by intra-wire scattering of Abelian quasiparticles (\ref{e/2scattering}). Then one can check that the composite electron operator is local only when
\begin{equation}\label{parityconstraint1}
\begin{cases}
\text{$l$ is even and $k$ is even,} &\text{for even $q$ (fermionic)}.\\
\text{$l$ is odd and $k$ is even,} &\text{for odd $q$ (bosonic)}.
\end{cases}
\end{equation}
This constrains the allowed filling fractions from (\ref{fillingfraction1}) to (\ref{fillingfraction2}). 

Let us now examine the tunneling term, by first separating the charge and neutral parts as follows,
\begin{equation}\label{e-tunnel}
\mathcal{H}_{1,j+1/2} = -t_1\Psi_{j+1,L}^{\dagger}\Psi_{j,R} +h.c. \equiv \mathcal{O}^{\sigma}\mathcal{O}^{\rho}+h.c.
\end{equation}
The charge part is simply $\mathcal{O}^{\rho}= e^{i\tilde{\theta}^{\rho,\uparrow}_{j+1/2}}$. As already discussed in Sec. \ref{sec4.2} , after gapping the charge sectors the link variable $\tilde{\theta}^{\rho,\uparrow}_{j+1/2}$ is already locked at integer multiples of $\pi$. We should then treat $\mathcal{O}^{\rho}$ as a pure number and absorb it into the coupling $t_1$. As for the neutral part, upon bosonizing $\gamma^{\uparrow}_r \sim \cos(\varphi^{\sigma}+r\theta^{\sigma})$, we obtain
\begin{equation}\label{oddneutral}
\begin{split}
&\mathcal{O}^{\sigma} =\sum_{\mathclap{\substack{r=R,L \\ n,m\in \text{ odd}}}} A^{r}_{n,m} \cos[\varphi^{\sigma}_j+r\varphi^{\sigma}_{j+1}-n\theta^{\sigma}_j-m\theta^{\sigma}_{j+1}]\\
&= \sum_{\mathclap{\substack{r=R,L \\ n,m\in \text{ even}}}} \tilde{A}^{r}_{n,m} \cos[\varphi^{\sigma}_j+r\varphi^{\sigma}_{j+1}+(\frac{p}{2}-n)\theta^{\sigma}_j+(\frac{p}{2}-m)\theta^{\sigma}_{j+1}].
\end{split}
\end{equation}
For latter convenience, in the second equality we have shifted the dummy indices $n, m$ by $p/2$. Then the specific term with $r=L=-1$ and $n=m=0$ corresponds to $\cos{\sqrt{\frac{p}{2}}(\bar{\phi}^{\sigma}_{j,R}-\bar{\phi}^{\sigma}_{j+1,L})}$, which is exactly what we need. However, $\mathcal{O}^{\sigma}$ contains many other terms as well which potentially complicate the physics. Fortunately, just as in the charge sector, we can introduce further inter-wire interaction to render $\cos{\sqrt{\frac{p}{2}}(\bar{\phi}^{\sigma}_{j,R}-\bar{\phi}^{\sigma}_{j+1,L})}$ as the only relevant term. Before we demonstrate that, let us also consider the tunneling term appropriate for even $p/2$, which despite being constructed differently, contain a similar expression for $\mathcal{O}^\sigma$. 

\subsubsection{p = 0 mod 4}
In this case we need to dress the electron operator with $\mathcal{V}_{4\sigma}$, which scatter a charge-$e/2q$ $\sigma$-particle in each layer from the right edge to the left edge. Combining (\ref{4sigmabosonized}) with (\ref{oddelectron}), we obtain
\begin{equation}\label{evenelectron}
\Psi_{j,r} \sim \Big(\sum_{\mathclap{n\in \text{ odd}}} B_n\cos n\theta^{\sigma}_j\Big) \gamma^{\uparrow}_{j,r} e^{i(\varphi^{\rho,\uparrow}_{j}+r\frac{l}{q}\theta^{\rho,\uparrow}_j+r\frac{k}{q}\theta^{\rho,\downarrow}_j)}.
\end{equation}
Due to the additional charge part of $\mathcal{V}_{4\sigma}$, the above composite electron operator is local only when
\begin{equation}\label{parityconstrait2}
\begin{cases}
\text{$l$ is odd and $k$ is odd,} &\text{for even $q$ (fermionic)}.\\
\text{$l$ is even and $k$ is odd,} &\text{for odd $q$ (bosonic)}.
\end{cases}
\end{equation}
Again, this constrains the allowed filling fractions from (\ref{fillingfraction1}) to (\ref{fillingfraction2}). Thus, all even-$p$ orbifold states share the same sequence of filling fractions which depends only on the fermionic/bosonic nature of the system.

Tunneling the composite electron appropriate for even $p/2$, the charge part of the tunneling term (\ref{e-tunnel}) is once again $\mathcal{O}^{\rho}= e^{i\tilde{\theta}^{\rho,\uparrow}_{j+1/2}}$, which can be absorbed into $t_1$ by gapping the charge sectors. What interests us more is the bonsonized representation of the neutral part
\begin{equation}\label{evenneutral}
\begin{split}
&\mathcal{O}^{\sigma} = \sum_{\mathclap{{\substack{r=R,L \\ n,m\in \text{ even}}}}} B^{r}_{n,m} \cos[\varphi^{\sigma}_j+r\varphi^{\sigma}_{j+1}-n\theta^{\sigma}_j-m\theta^{\sigma}_{j+1}]\\
&= \sum_{\mathclap{{\substack{r=R,L \\ n,m\in \text{ even}}}}} \tilde{B}^{r}_{n,m} \cos[\varphi^{\sigma}_j+r\varphi^{\sigma}_{j+1}+(\frac{p}{2}-n)\theta^{\sigma}_j+(\frac{p}{2}-m)\theta^{\sigma}_{j+1}].
\end{split}
\end{equation}
Once again, the inter-wire tunneling term that we need appears in the second equality above when $r=L=-1$ and $n=m=0$.

Comparing (\ref{oddneutral}) and (\ref{evenneutral}), we see that although different constructions of the tunneling operator are adopted for different parities of $p/2$, the same bosonized form is reached. As we will see below, the exact values of the coefficients $\tilde{A}$ and $\tilde{B}$ are immaterial to us, so let us summarize the above two cases and express in terms of the neutral chiral fields (\ref{neutralchiralfields}) as follows,
\begin{equation}
\begin{split}
\mathcal{O}^{\sigma} = \sum_{\substack{r=R,L \\ n,m\in \mathbb{Z}}} &C^{r}_{n,m} \cos\sqrt{\frac{2}{p}}\;[\;n\bar{\phi}^{\sigma}_{j,L} + (\frac{p}{2}-n) \bar{\phi}^{\sigma}_{j,R} \\
&+(r\frac{p}{2}-m)\bar{\phi}^{\sigma}_{j+1,L}+ m\bar{\phi}^{\sigma}_{j+1,R}\;].
\end{split}
\end{equation}
For the tunneling operator to gap the neutral sector, it is required that $C^{L}_{0,0}$ flows to strong coupling exclusively while all other $C^{r}_{n,m}$ vanish at low energy. This is achieved by an inter-wire scattering to be discussed next. 
\subsection{\label{sec5.2}Gapping the neutral sector}
Consider an inter-wire interaction analogous to the one in (\ref{chargeinterwire}) so that the array of wires has the following bulk Hamiltonian in the neutral sector,
 \begin{equation}\label{neutralinterwire}
 \begin{split}
\tilde{\mathcal{H}}^{\sigma}_{0,\text{bulk}} =\frac{\bar{v}_\sigma}{2\pi} \sum_{j=1}^{N-1}[(\partial_x \bar{\phi}^{\sigma}_{j,R})^2 
&+ 2\lambda_\sigma(\partial_x\bar{\phi}^{\sigma}_{j,R})(\partial_x \bar{\phi}^{\sigma}_{j+1,L})\\ &+ (\partial_x \bar{\phi}^{\sigma}_{j+1,L})^2].
\end{split}
 \end{equation}
 
\noindent While one might have noticed that the additional inter-wire term $\lambda_\sigma(\partial_x\bar{\phi}^{\sigma}_{j,R})(\partial_x \bar{\phi}^{\sigma}_{j+1,L})$ is not an allowed local operator, one can actually construct a local operator with this term in the neutral sector by attaching an appropriate charge-sector factor (see Appendix \ref{secappendixa}). Upon gapping the charge sector, the charge variable is locked so that it can be absorbed into the coupling constant $\lambda_\sigma$. It is in this sense that (\ref{neutralinterwire}) provides a well-defined local Hamiltonian for our analysis.
 
Proceeding with $\tilde{\mathcal{H}}^{\sigma}_{0,\text{bulk}}$, it can be diagonalized by the following link variables,
\begin{subequations}
\begin{align}
\widetilde{\theta}^{\sigma}_{j+1/2} &= (\bar{\phi}^{\sigma}_{j,R}-\bar{\phi}^{\sigma}_{j+1,L})/2,\\
\widetilde{\varphi}^{\sigma}_{j+1/2} &=(\bar{\phi}^{\sigma}_{j,R}+\bar{\phi}^{\sigma}_{j+1,L})/2,
\end{align}
\end{subequations}
which obey
\begin{equation}
\begin{split}
[\tilde{\theta}_{j+1/2}^{\sigma}, \tilde{\theta}_{j'+1/2}^{\sigma}] =[\tilde{\varphi}_{j+1/2}^{\sigma} , \tilde{\varphi}_{j'+1/2}^{\sigma} ]=0, \\
[\partial_x \tilde{\theta}_{j+1/2}^{\sigma}, \tilde{\varphi}_{j'+1/2}^{\sigma}] =i\pi \delta_{j,j'}\delta_{x,x'}.
\end{split}
\end{equation}
Then, analogous to (\ref{linkHamiltonian}), the neutral sector Hamiltonian can be written as
\begin{equation}
\tilde{\mathcal{H}}^{\sigma}_{0,\text{bulk}} =\frac{\tilde{v}_\sigma}{\pi} \sum_{j=1}^{N-1} [\frac{1}{g_\sigma}(\partial_x\tilde{\theta}^{\sigma}_{j+1/2})^2+g_\sigma(\partial_x\tilde{\varphi}^{\sigma}_{j+1/2})^2],
\end{equation}
where $g_\sigma =\sqrt{(1+\lambda_\sigma)/(1-\lambda_\sigma)}$ and $\tilde{v}_\sigma = \bar{v}_\sigma \sqrt{1-\lambda_\sigma^2}$. Consequently, by tuning $\lambda_\sigma \rightarrow -1^-$, the scaling dimension of $e^{i n\widetilde{\varphi}^\sigma}$ will diverge, while that of $e^{in\widetilde{\theta}^\sigma}$ will vanish.

Equipped with this observation, let us re-write $\mathcal{O}^{\sigma}$ in terms of the link variables, so that the scaling dimension of each term can be easily deduced. We have
\begin{widetext}
\begin{equation}
\mathcal{O}^{\sigma} = \sum_{\substack{r=R,L \\ n,m\in \mathbb{Z}}} C^{r}_{n,m} \cos\sqrt{\frac{2}{p}}\Big\{n(\tilde{\varphi}^\sigma_{j-1/2}-\tilde{\theta}^\sigma_{j-1/2})+[(1+r)\frac{p}{2}-n-m]\;\tilde{\varphi}^\sigma_{j+1/2}+[(1-r)\frac{p}{2}-n+m]\;\tilde{\theta}^\sigma_{j+1/2}+m(\tilde{\varphi}^\sigma_{j+3/2}+\tilde{\theta}^\sigma_{j+3/2})\Big\}.
\end{equation}
\end{widetext}
Though the above expression looks complicated, our observation above suggests that any cosine terms containing $\tilde{\varphi}^{\sigma}$ would be rendered irrelevant at low energy. It is clear that the only relevant term, whose scaling dimension can be engineered to be arbitrarily small, corresponds to $C^{L}_{0,0,}$.

As such, considering the coupled wire model described by
\begin{equation}
\mathcal{H}^\sigma_{\text{bulk}} = \tilde{\mathcal{H}}^\sigma_{0,\text{bulk}} + \sum_{j=1}^{N-1} \mathcal{H}_{1, j+1/2}\;, 
\end{equation}
the bulk neutral sector is gapped following the above arguments. $C^{L}_{0,0,}$ flows exclusively to strong coupling and leads to the locking of bulk variables $\tilde{\theta}^\sigma_{j+1/2}$. There are two edge modes, $\bar{\phi}^{\sigma}_{1,L}$ and $\bar{\phi}^\sigma_{N,R}$, being decoupled from the bulk and thus are left fluctuating freely. As we have seen in Sec. \ref{sec3.2}, these gapless edge modes are described by the orbifold conformal field theory at radius $R_{\rm orbifold}=\sqrt{p/2}$ with even integer $p$. This completes the coupled wire construction for the $Z_2 \times Z_2$ orbifold quantum Hall states. 

\section{\label{sec7}Spectrum of Quasiparticles}
In this section, we briefly discuss the structure of quasiparticles in the $Z_2 \times Z_2$ orbifold states just constructed. We are not going to derive from our microscopic theory the topological $\mathcal{S}$ matrix, which already exists in the literature \citep{2} and does not depend on the exact implementation of the coupled wire model. Instead, we will focus on characterizing quasiparticle sectors by the amount of charge to which one can attach, which is model-dependent. This will determine the conformal dimension $h_\alpha$ of the allowed quasiparticles. Together with the central charge of our model, which is $c=3$ (as there are two charge sectors with $c=1$ each and one neutral sector with $c=1$), we can obtain the topological $\mathcal{T}$ matrix
\begin{equation}
\mathcal{T}_{\alpha\beta} = \delta_{\alpha\beta} e^{2\pi i (h_\alpha-c/24)}
\end{equation}
with $\alpha,\beta$ labeling quasiparticle sectors. The $\mathcal{S}$ and $\mathcal{T}$ matrices together encode all the topological properties, including braiding statistics, of the bulk quasiparticles. 

Our reasoning relies on the edge theory, which is described by the chiral charge fields $\bar{\phi}^{\rho,\uparrow/\downarrow}_{R/L}$ and the chiral neutral fields $\bar{\phi}^{\sigma}_{R/L}$. Any allowed quasiparticle operator must be a combination of primary fields in the charge and neutral sectors, so that back-scattering of a quasiparticle is represented by
\begin{equation}
\Omega_R^\sigma \Omega_L^\sigma\; e^{i[Q^\uparrow(\bar{\phi}^{\rho,\uparrow}_R-\bar{\phi}^{\rho,\uparrow}_L)+Q^\downarrow(\bar{\phi}^{\rho,\downarrow}_R-\bar{\phi}^{\rho,\downarrow}_L)]},
\end{equation}
where $2eQ^{\uparrow/\downarrow}$ is the charge carried by the quasiparticle in the $\uparrow/\downarrow - $layer. Our task is to determine what combinations of $\Omega^\sigma$ and $(Q^\uparrow,Q^\downarrow)$ are allowed by locality, or in other words, expressible as a product of local operators of the original MR $\times$ MR state. 

Quasiparticles that are trivial in the neutral sector can be obtained by considering the following operator already introduced in Sec. \ref{sec3.1.2},
\begin{equation}
\begin{split}
(\mathcal{V}_1^{a})^\frac{\chi^a}{2} &\sim e^{\frac{i\chi^a}{q}\theta^{\rho,a}} \\
&= \exp \{\frac{i\chi^a}{4(k^2-l^2)}[k(\bar{\phi}^{\rho,\bar{a}}_R-\bar{\phi}^{\rho,\bar{a}}_L)-l(\bar{\phi}^{\rho,a}_R-\bar{\phi}^{\rho,a}_L)]\}.
\end{split}
\end{equation}
Here $a=\uparrow/\downarrow$ and $\bar{a}=\downarrow/\uparrow$, and the change of variables in (\ref{newchargechiral}) has been used. The above operator is local as long as $\chi^a$ is an integer multiple of 2. In the context of orbifold states, it represents a transfer of quasiparticle from right to left, with net charge (adding up the electric charge contributed by the two charge sectors)
\begin{equation}
e\cdot\frac{\chi^a}{2}(\frac{k}{k^2-l^2}-\frac{l}{k^2-l^2})= \frac{e\chi^a}{2(k+l)} = \frac{\nu e}{4}\chi^a .
\end{equation}
The last equality is obtained using (\ref{fillingfraction1}). Such a quasiparticle is trivial in the neutral sector, and can be attached to any quasiparticles without modifying their topological sectors. Motivated by this observation, we express the charge of a generic quasiparticle (in unit of $2e$) as follows, 
\begin{equation}\label{layercharge}
(Q^\uparrow,Q^\downarrow) = \frac{\chi^\uparrow}{4(k^2-l^2)}(-l,\;k)+\frac{\chi^\downarrow}{4(k^2-l^2)}(k,\;-l).
\end{equation}
By defining $\bm{Q} = (Q^\uparrow,Q^\downarrow)^T$ and $\bm{\chi}=(\chi^\uparrow,\chi^\downarrow)^T$, the above expression can be compactly written as
\begin{equation}
\bm{Q}  = K^{-1}\bm{\chi},
\end{equation}
which is in the familiar form for a bilayer Halperin state characterized by the $K$-matrix in (\ref{Kmatrix2}). As we will see next, the quasiparticle sectors in the $Z_2 \times Z_2$ orbifold states are classified by $\bm{\chi}$ mod 2. 

The charge assignment can actually be inferred from the fusion rules in the orbifold theory. According to (\ref{nonAbelianfusion}), quasiparticles in the $1,\phi_p^{1,2}$ and $\phi_{k=\rm even}$ sectors should have the same charge, up to modification by quasiparticles with a trivial neutral sector, and thus are all neutral in this sense. Then (\ref{Z2fusion}) further suggests that the $j$ sector is neutral as well. The fusion rules also imply that the twisted sectors should carry half the charge of the untwisted sectors. Finally, by requiring that the charge assignment be compatible with the known spectrum of MR $\times$ MR, which is the starting point of our model, we are left with the following classification,
\begin{equation}
\begin{alignedat}{3}
(\chi^\uparrow,\chi^\downarrow)&=(0,0) \text{ mod }2,\quad \alpha&&=1,\;j,\;\phi_p^{1,2},\;\phi_{k=\rm even}.\\
(\chi^\uparrow,\chi^\downarrow)&=(1,1) \text{ mod }2,\quad \alpha&&=\phi_{k=\rm odd}.\\
(\chi^\uparrow,\chi^\downarrow)&=(0,1) \text{ mod }2,\quad \alpha&&= \sigma_1,\;\tau_1.\\
(\chi^\uparrow,\chi^\downarrow)&=(1,0) \text{ mod }2,\quad \alpha&&= \sigma_2,\;\tau_2.
\end{alignedat}
\end{equation}
With this result, the conformal dimension of quasiparticles $h_\alpha$, and subsequently the topological $\mathcal{T}$ matrix, can be determined. 

From the experimental perspective, it is also interesting to characterize the spectrum by the net electric charge of quasiparticles, which is easier to measure than the exchange statistics. In terms of the net charge
\begin{equation}
Q \equiv 2e(Q^\uparrow+Q^\downarrow) =\frac{\nu e}{4}\chi_{\rm total} ,
\end{equation}
where we have defined $\chi_{\rm total} \equiv \chi^\uparrow+\chi^\downarrow$, we are left with the following charge spectrum,
\begin{equation}
\begin{split}
\chi_{\rm total} &=0 \text{ mod 2}\;:\quad \alpha=1,\;\;j,\;\;\phi_p^{1,2},\;\;\phi_{k}.\\
\chi_{\rm total} &=1 \text{ mod 2}\;: \quad \alpha=\sigma_{1,2},\;\;\tau_{1,2}.
\end{split}
\end{equation}

\section{\label{sec6}Discussion}
In this paper we have presented a microscopic construction for the $Z_2 \times Z_2$ orbifold quantum Hall states by coupling wires of MR$\times$MR-bilayer. These are non-Abelian states with the neutral sector characterized by the $c=1$ orbifold conformal field theory at radius $R_{\rm orbifold} =\sqrt{p/2}$ with even integers $p$. The spectrum of quasiparticles have been studied, where quasiparticles can be classified into two groups according to their electric charge. These orbifold states are shown to exist at the following filling factors
\begin{equation}
\nu=\begin{cases}
1/n, &\text{(for fermions)}\\
2/(2n+1), &\text{(for bosons)}
\end{cases}
\end{equation}
with $n\in\mathbb{Z}$, for all even integers $p$. We would like to point out that the even-$p$ orbifold states constructed here, although motivated by the proposal in Ref.\citep{5} and thus share certain similarities in places like the fusion rules, are not identical to the orbifold states introduced by Barkeshli and Wen. Particularly, the even-$p$ orbifold states discussed there occur at filling
\begin{equation}\label{BWfilling}
\nu_{\rm BM}=\begin{cases}
1/2n, &\text{(for $p=2$ mod $4$, fermions)}\\
1/(2n+1), &\text{(for $p=0$ mod $4$, fermions)}\\
1/(2n+1), &\text{(for $p=2$ mod $4$, bosons)}\\
1/2n, &\text{(for $p=0$ mod $4$, bosons)}\\
\end{cases}
\end{equation}
with $n\in\mathbb{Z}$. Therefore, with the underlying electrons being fermionic, orbifold states constructed in this paper actually occur at a more extensive set of filling factors than the ones introduced by Barkeshli and Wen. Another important difference lies in the central charge, as our model has $c=3$ (with two $c=1$ charge sectors and one $c=1$ orbifold neutral sector) while the model of Barkeshli and Wen has $c=2$. The $p=2$ case in our model describes two copies of Moore-Read Pfaffian states, while in their case it is a Pfaffian state with an extra copy of Ising model.

Hence, one possible extension of this work would be to consider a coupled wire model with wires of MR$\times$Ising-bilayer. If the bulk of this model can be gapped by local interactions, one can obtain orbifold states with central charge $c=2$. Following similar arguments as those presented in this paper, one can argue that such orbifold states would occur at filling $\nu=1/l$, where $l$ has to satisfy the parity constraints in either (\ref{parityconstraint1}) or (\ref{parityconstrait2}) depending on $p$ mod $4$. In this way, one can recover the sequence of filling factors in (\ref{BWfilling}), as proposed by Barkeshli and Wen. However, it is a non-trivial task to introduce local interactions that sew together the MR$\times$Ising wires so as to obtain a gapped bulk and gapless edges described by the orbifold conformal field theory. Certain progress can be made by considering a Gross-Neveu-like interaction for two pairs of Majorana fields from consecutive wires. This is a local interaction that can sew an array of Ising wires together and give rise to a gapped two-dimensional bulk with a pair of gapless Ising edges, as demonstrated in Ref.\citep{16}. It is however unclear how to generalize this procedure to obtain orbifold states for $p>2$. Alternatively, one may consider an explicit implementation of the Ising model, through which local operators, similar to the ones adopted here for intra/inter-wire interactions, can be identified. We will leave this for future work. 

\begin{acknowledgments}
The authors would like to thank Ady Stern for helpful discussions. This work is in part supported by the Croucher Scholarship for Doctoral Study from the Croucher Foundation (PMT), grant EP/S020527/1 from EPSRC (YH) and a Simons Investigator grant from the Simons Foundation (CLK). 
\end{acknowledgments}

\appendix
\renewcommand{\theequation}{A.\arabic{equation}}
\section{\label{secappendixa}Repairing the Neutral Sector}
In Sec. \ref{sec5.2}, we argue about how an inter-wire interaction of the form $\lambda_\sigma(\partial_x\bar{\phi}^{\sigma}_{j,R})(\partial_x \bar{\phi}^{\sigma}_{j+1,L})$ can render only the desired tunneling term $C^{L}_{0,0}$ relevant at low energy and lead to the gapping of neutral sector. However, it is not trivial to see why this term is actually allowed to appear in a Hamiltonian. Here, we will settle this issue by explicitly constructing a local interaction with such a form in the neutral part, and with a charge part that can be condensed by gapping charge sectors.

Let us begin with interactions that can be easily seen to be local, such as
\begin{subequations}
\begin{align}
\Theta_R &= \Theta^\sigma_R \Theta^\rho= \gamma^\uparrow_{R}\gamma^\downarrow_{R}e^{\frac{i}{2}\bar{\phi}^{\rho, \uparrow}_{R}}e^{\frac{i}{2}\bar{\phi}^{\rho, \downarrow}_{R}},\\
\Theta_L &=  \Theta^\sigma_L \Theta^\rho= \gamma^\uparrow_{L}\gamma^\downarrow_{L}e^{\frac{i}{2}\bar{\phi}^{\rho, \uparrow}_{R}}e^{\frac{i}{2}\bar{\phi}^{\rho, \downarrow}_{R}}.
\end{align}
\end{subequations}
We first focus on operators in a single wire so the wire label $j$ is suppressed for simplicity. From the definition of the charged chiral fields in (\ref{newchargechiral}), the charge part of the above operator can be written in two equivalent ways as
\begin{equation}
\Theta^\rho \sim 
\begin{cases}
e^{i\phi^{\rho,\uparrow}_R}e^{i\phi^{\rho,\downarrow}_R}e^{i\frac{l+k-q}{q}(\theta^{\rho,\uparrow}+\theta^{\rho, \downarrow})}, \quad\text{for }\Theta_R.\\
e^{i\phi^{\rho,\uparrow}_L}e^{i\phi^{\rho,\downarrow}_L}e^{i\frac{l+k+q}{q}(\theta^{\rho,\uparrow}+\theta^{\rho, \downarrow})}, \quad\text{for }\Theta_L.
\end{cases}
\end{equation}
It is then clear that we can interpret $\Theta_{R/L}$ as annihilation of a MR edge electron and $(l+k\pm q)/2$ scatterings of charge-$e/q$ Abelian quasiparticles in each layer. Following the parity requirements in (\ref{parityconstraint1}) or (\ref{parityconstrait2}), $(l+k\pm q)/2$ is indeed an integer and hence $\Theta_{R/L}$ is local.

We then bosonize $i\gamma^{\uparrow}_R\gamma^{\downarrow}_R \sim \partial_x \phi^{\sigma}_R$ and $i\gamma^{\uparrow}_L\gamma^{\downarrow}_L \sim \partial_x \phi^{\sigma}_L$. It can be seen from (\ref{neutralchiralfields}) that $\bar{\phi}^\sigma_{R}$ is a linear combination of $\phi^{\sigma}_R$ and $\phi^{\sigma}_L$, and hence by taking linear combination of two local operators $\Theta_R$ and $\Theta_L$, we conclude that the following operator is also local,
\begin{equation}
\Xi_{R} \sim \partial_x \bar{\phi}^\sigma_R\;e^{\frac{i}{2}\bar{\phi}^{\rho, \uparrow}_{R}}e^{\frac{i}{2}\bar{\phi}^{\rho, \downarrow}_{R}}.
\end{equation}
Analogously, one can argue that 
\begin{equation}
\Xi_{L} \sim \partial_x \bar{\phi}^\sigma_L\;e^{\frac{i}{2}\bar{\phi}^{\rho, \uparrow}_{L}}e^{\frac{i}{2}\bar{\phi}^{\rho, \downarrow}_{L}}
\end{equation}
is an allowed local operator. Now we can put back the wire-labels and consider a product of the above two operators (on successive wires),
\begin{equation}
\begin{split}
&\Xi_{j+1, L}^\dagger \Xi_{j,R} + h.c.  \\
&\sim (\partial_x \bar{\phi}^\sigma_{j,R})(\partial_x \bar{\phi}^\sigma_{j+1,L}) \cos(\tilde{\theta}^{\rho,\uparrow}_{j+1/2}+\tilde{\theta}^{\rho,\downarrow}_{j+1/2}).
\end{split}
\end{equation}
Upon gapping the charge sectors, link variables $\tilde{\theta}^{\rho,\uparrow/\downarrow}$ are pinned at integer multiples of $\pi$. The above cosine term then simply takes value of $\pm1$, and hence can be safely discarded. This justifies our argument in the main text, where we treat $(\partial_x \bar{\phi}^\sigma_{j,R})(\partial_x \bar{\phi}^\sigma_{j+1,L})$ as a local operator and allow it to appear in the bulk neutral sector Hamiltonian in (\ref{neutralinterwire}).

\bibliographystyle{apsrev4-1.bst}
\bibliography{orbifoldQH}

\end{document}